\documentclass[10pt]{article}

\renewcommand{\baselinestretch}{1.0}

\usepackage{a4wide}

\usepackage{epsfig}

\usepackage{fancyheadings}
\usepackage{cite}
\usepackage{amsmath}
\usepackage{amssymb}

\begin{document}
%

%
%

\def\RS{Randall-Sundrum }
\def\KK{KK }
\def\lap{\square}
\def\grad{\nabla}

\def\olap{{}^{\scriptscriptstyle 0} \square}
\def\indlap{{}^{\scriptscriptstyle I} \square}
\def\ograd{{}^{\scriptscriptstyle 0} \nabla}
\def\indgrad{{}^{\scriptscriptstyle I} \nabla}
\def\g0{{}^{\scriptscriptstyle 0} g}
\def\gind{{}^{\scriptscriptstyle I} g}
\def\R0{{}^{\scriptscriptstyle 0} R}
\def\G0{{}^{\scriptscriptstyle 0} G}
\def\indT{{}^{\scriptscriptstyle I} T}
\def\indR{{}^{\scriptscriptstyle I} R}
\def\indG{{}^{\scriptscriptstyle I} G}

\def\Mlap{{}^{\scriptscriptstyle \eta} \square}
\def\Mg{{}^{\scriptscriptstyle \eta} g}
\def\Mh{{}^{\scriptscriptstyle \eta} h}

\def\egrad{\sqrt{\epsilon} \, \nabla}

\def\bx{\bar{x}}
\def\bz{\bar{z}}

\newcommand{\dz}[1]{\partial_z #1}
\newcommand{\ddz}[1]{\partial_z^2 #1}
\newcommand{\dbz}[1]{\bar{\partial_{z}} #1}

\def\figuremode{\small}

%
%

\title{\bf Strong Brane Gravity and the Radion at Low Energies }

\author{{\bf Toby Wiseman}\thanks{e-mail: {\tt
      T.A.J.Wiseman@damtp.cam.ac.uk}} \\ \\
  Department of Applied Mathematics and Theoretical Physics,\\
  Center for Mathematical Sciences,\\
  Wilberforce Road,\\
  Cambridge CB3 0WA, UK}

\date{January 16, 2002}

\maketitle

%
\begin{abstract}
%
  
  For the 2-brane Randall-Sundrum model, we calculate the bulk
  geometry for strong gravity, in the low matter density regime, for
  slowly varying matter sources. This is relevant for astrophysical or
  cosmological applications.  The warped compactification means the
  radion can not be written as a homogeneous mode in the orbifold
  coordinate, and we introduce it by extending the coordinate patch
  approach of the linear theory to the non-linear case.  The negative
  tension brane is taken to be in vacuum. For conformally invariant
  matter on the positive tension brane, we solve the bulk geometry as
  a derivative expansion, formally summing the `Kaluza-Klein'
  contributions to all orders.  For general matter we compute the
  Einstein equations to leading order, finding a scalar-tensor theory
  with $\omega(\Psi) \propto \Psi / (1 - \Psi)$, and geometrically
  interpret the radion.  We comment that this radion scalar may become
  large in the context of strong gravity with low density matter.
  Equations of state allowing $(\rho - 3 P)$ to be negative, can
  exhibit behavior where the matter decreases the distance between the
  2 branes, which we illustrate numerically for static star solutions
  using an incompressible fluid.  For increasing stellar density, the
  branes become close before the upper mass limit, but after violation
  of the dominant energy condition.  This raises the interesting
  question of whether astrophysically reasonable matter, and initial
  data, could cause branes to collide at low energy, such as in
  dynamical collapse.

%
\end{abstract}
%

\vspace{5.0cm}
\begin{flushright}
DAMTP-2002-4 \\
hep-th/0201127
\end{flushright}

\newpage

%
\section{Introduction}
%

Much progress has been made in understanding the long range
gravitational response of branes at orbifold fixed planes, to
localized matter. The Randall-Sundrum models, with one or two branes
\cite{Randall:1999ee,Randall:1999vf,Lykken:1999nb} simply have gravity
and a cosmological constant in the bulk, making the linear problem
very tractable. Linear calculations
\cite{Garriga:1999yh,Giddings:2000mu,Tanaka:2000zv,Sasaki:1999mi} show
that in the one brane case, long range $4$-dimensional gravity is
recovered on the brane. Little is known of the general non-linear
behavior \cite{Emparan:1999wa,Emparan:2001ce}.  There is no mass gap,
and thus the non-linear problem is essentially a 5-dimensional one,
even for long wavelength sources.  Second order perturbation theory
\cite{Kudoh:2001wb,Giannakis:2000zx}, and fully non-linear studies
\cite{Wiseman:2001,Gregory:2001xu} are again consistent with
recovering usual 4-dimensional strong gravity.  In particular
\cite{Wiseman:2001} shows that non-perturbative phenomena, such as the
upper mass limit for static stars, extend smoothly from large to small
objects, whose characteristic sizes are taken relative to the AdS
length.

For the two brane case, linear theory \cite{Garriga:1999yh} found that
the effective gravity is Brans-Dicke, for an observer on the positive
tension brane, and a vacuum negative tension brane.  For reasonable
brane separations phenomenologically acceptable gravity is recovered.
For observers on the negative tension brane it was found the response
was incompatible with observation for any brane separation, although
stabilizing the distance between the branes does allow one to recover
standard 4-dimensional gravity
\cite{Goldberger:1999uk,Goldberger:1999un,Goldberger:1999wh,Csaki:1999mp,Tanaka:2000er,Mukohyama:2001ks,Kudoh:2001kz}.
In this paper we consider an observer on the positive tension brane,
the negative tension brane to be in vacuum, and the orbifold radius to
be unstabilized. This allows us to study the dynamics of the radion in
strong gravity. Of course, our methods may be extended to include the
stabilized case too. Although moduli have previously been assumed to be
fixed at late times, a crucial feature of the recent cyclic Ekpyrotic
scenario \cite{Steinhardt:2001st,Steinhardt:2001vw}, is that the
radius of the orbifold is not stabilized, the cyclic nature of this
model being such that the separation always remains finite.

The two brane strong gravity case appears more tractable than for one
brane, as the linear theory allows a Kaluza-Klein style reduction of
the propagator \cite{Garriga:1999yh}. In Kaluza-Klein
compactifications, only the homogeneous zero modes are excited on long
wavelengths, and the matter is thought to comprise fields over the
whole internal space.  However, for matter to be supported on the
orbifold branes there must be modes excited which are not homogeneous
in the extra dimension. Thus one cannot simply write down a non-linear
ansatz for the metric, as is familiar from Kaluza-Klein
compactifications.

In previous work on the Horava-Witten compactification
\cite{Horava:1996ma}, the low energy, long range effective theory was
constructed using consistent orbifold reductions
\cite{Lalak:1998ti,Lukas:1998fg,Lukas:1998tt,Lukas:1998ew,Lukas:1999yn}.
The methods developed treated the homogeneous component of the metric
as a background, and inhomogeneous perturbations about it as the
contribution of the massive `Kaluza-Klein' modes. The metric was then
solved to leading order in a derivative expansion. Strong gravity was
not discussed in these works except when considering inflation
\cite{Lukas:1999yn}. In order to use these constructions, all the zero
modes of the orbifold must be homogeneous. One important result of the
warped compactification, is that whilst the graviton modes are
homogeneous, the radion zero mode is not \cite{Charmousis:1999rg}.
This appears to be a general feature of warped models
\cite{Gen:2000nu,Chacko:2001em,Kogan:2001qx}, and one cannot directly
use these Horava-Witten methods.

The aim of this paper is to apply the orbifold reduction to the warped
Randall-Sundrum model, and then examine the non-linear behavior of
the radion. We begin in section \ref{sec:scalar} by illustrating the
reduction proposed in
\cite{Lalak:1998ti,Lukas:1998fg,Lukas:1998tt,Lukas:1998ew,Lukas:1999yn}
with a field theory example and discuss its application to strong
gravity. In order to apply the method we must find a way to include
the inhomogeneous radion zero mode. Instead of explicitly using an
ansatz including the radion, we include it by `deflecting' the brane
relative to a non-linear extension of the `Randall-Sundrum gauge'
\cite{Garriga:1999yh,Giddings:2000mu}. In sections \ref{sec:metric}
and \ref{sec:conf} we solve the bulk metric for conformally invariant,
low energy, long wavelength matter on the positive tension brane. For
conformally invariant matter, this coordinate system is Gaussian
normal to the brane. The bulk geometry is constructed using a
derivative expansion, which is formally summed to all orders.

In section \ref{sec:nonconf}, we extend the linear analysis to the
non-linear, for general low density, long wavelength matter, allowing
the brane to become deflected relative to this coordinate system.
This scalar deflection becomes the radion, and is treated
non-linearly.  This is crucial, as this field takes values roughly of
order the Newtonian potential, for static systems.  Therefore, the
case of strong gravity is exactly where the radion must be treated
non-linearly.  We calculate a local action for the effective induced
Einstein equations, to leading order in the derivative expansion.  The
resulting theory is a scalar-tensor model, which reduces to
Brans-Dicke in the linear approximation. This is the first non-linear,
covariant derivation of the effective action.  Considering the zero
modes \cite{Chiba:2000rr} does not tell one the conformal metric that
matter couples to. Whilst cosmological solutions such as
\cite{Csaki:1999mp,Goldberger:1999un,Binetruy:1999hy,Binetruy:2001tc}
may derive radion dynamics, this only applies for a homogeneous
radion, and thus is not a covariant derivation.  Indeed the examples
\cite{Csaki:1999mp,Goldberger:1999un} illustrate this point well.  The
form of the metric chosen, whilst coincidentally reproducing the
correct effective action \cite{Chiba:2000rr}, does not solve the
correct linearized equations \cite{Charmousis:1999rg}, and thus only
applies to the case of cosmological symmetry.

We discuss in detail the validity of the approximation for considering
low density, long wavelength strong brane gravity. Terms that are
quadratic in the energy density are neglected in the orbifold
reduction method, consistent with the low energy approximation. The
characteristic length scale of the matter must be large compared to
the compactification scale for the derivative expansion to be valid.
For strong gravity, the approximation then holds provided the
4-dimensional induced brane curvature invariants are small, compared
to the compactification scales. Thus this method will not allow a
global solution to a black hole geometry, or any other spacetimes with
curvature singularities.  However, it will apply to all other cases of
strong gravity, such as static relativistic stars, an example being
neutron stars, or dynamical non-linear systems, such as binary neutron
star systems, or collapse of matter up to the point where curvatures
become singular.

Having pointed out that the radion may take large values for strong
gravity configurations, we illustrate this using the example of
relativistic stars.  Previous work
\cite{Germani:2001du,Deruelle:2001fb} has considered stars on branes
using the projection formalism of \cite{Shiromizu:1999wj}, allowing
the quadratic stress tensor corrections to be calculated by making
ansatzes on the bulk geometry. However, these corrections are assumed
to be negligible in our low energy density assumption, and it is the
bulk geometric corrections which are relevant, and cannot be
calculated in the projection approach. In \cite{Wiseman:2001} the full
bulk solutions were numerically constructed for the one brane
Randall-Sundrum case.  Using the methods here, we are able to
analytically construct the geometry for large stars in the 2-brane
case. In section \ref{sec:star}, using the leading order action, we
numerically consider the static relativistic star, for incompressible
fluid matter.  The linear theory shows that for perturbative stars
with positive density the branes are deflected apart.  However, it
indicates that if $\rho - 3 P$ becomes negative the opposite could
occur. Now understanding the radion and bulk geometry non-linearly, we
consider this, finding that as non-linear effects become important the
branes do indeed become closer.  Furthermore, the branes appear to
meet \emph{before} the upper mass limit is reached. This will occur
for any brane separation.  However, for phenomenologically acceptable
separations \cite{Garriga:1999yh}, we find it does not occur before
the dominant energy condition is violated. Neutron stars are believed
to have polytropic equations of state that do not support negative
$\rho - 3 P$. It then remains a very interesting, and tractable $1+1$
problem to understand whether dynamical systems may cause branes to
collide, for realistic matter and initial data. We then have the
possibility that physical matter at low energies and curvatures,
compared to compactification scales, might cause brane collisions
requiring a Planck energy physics description.

%
\section{Orbifold Reduction: A Field Theory Example}
\label{sec:scalar}
%

We illustrate the method of consistent orbifold reduction using a
simple scalar field theory. The method was originally used in the
context of Horava-Witten reductions
\cite{Lalak:1998ti,Lukas:1998fg,Lukas:1998tt,Lukas:1998ew,Lukas:1999yn}.
We take the scalar equation,
\begin{equation}
\lap_d \Psi + (\partial_\mu \Psi)^2 + \partial_y^2 \Psi + 
(\partial_y \Psi)^2 = 0
\end{equation}
in a Minkowski $(d+1)$-dimensional bulk with signature $-,+,+,\ldots$
and coordinates $x^\mu, y$, with $\mu = 0, 1, \ldots (d-1)$. We
consider a finite range for $y$, choosing units such that $0 \le y \le
1$.  The operator $\lap_d$ is the Laplacian on $d$-dimensional
Minkowski space formed from $x^\mu$.  The analogy of a Kaluza-Klein
compactifications is to use periodic boundary conditions in $y$,
identifying the field at $y = 0$ and $y = 1$.  For orbifold brane
compactifications the boundary conditions to consider are Neumann,
with the field gradient $\partial_y \Psi = \rho(x)$ at $y = 0$ and
being zero gradient at $y = 1$. This will later correspond to the
case of matter on the positive tension brane, and a vacuum negative
tension brane. The zero modes of the system, with $\rho(x) = 0$ are
simply,
\begin{equation}
\Psi(x,y) = \Psi(x), \qquad \mbox{where,} \quad \lap_d \Psi +
(\partial_\mu \Psi)^2 = 0
\end{equation}
Having no $y$ gradient this induces no $\rho(x)$ on the boundary, and
in brane language, is analogous to the zero mode ansatz discussed in
\cite{Charmousis:1999rg}. Of course this solution also solves the
Kaluza-Klein periodic boundary conditions, as these are identical to
the Neumann ones if $\rho(x) = 0$. Then the full non-linear equations
are solved in terms of a lower dimensional system. However, for
non-trivial $\rho(x)$ the solution cannot be independent of $y$, and
the question is whether one can reduce the problem to a
$d$-dimensional one. 

If $\mid \rho \mid << 1$ we can linearize giving $\lap_{d} \Psi +
\partial_y^2 \Psi = 0$, whose solutions can obviously be found
exactly. For slowly varying matter, we may use a derivative expansion
in $\lap_d$ as,
\begin{equation}
\Psi(x,y) = \left( \frac{1}{\lap_d} + \left( - \frac{1}{3} + y -
    \frac{1}{2} y^2 \right) + O( \lap_d ) \right)
\rho(x)
\label{eq:toyexp}
\end{equation}
where we implicitly assume that the $d$-dimensional boundary
conditions, relevant for the particular problem, are taken into
account when evaluating the inverse Laplacian. It is these boundary
conditions that specify the zero mode component in the solution
$\Psi$. However we now see that for large enough sources, of
characteristic spacetime scale $L$ and density $\rho$, the leading
term will be large for $\rho L^2 \sim 1$. Then the non-linear terms in
the original equation cannot be neglected. For low density matter,
$\mid \rho \mid << 1$, we require $L >> 1$ for such strong
gravitational effects, and then we expect higher terms in the
derivative expansion to become smaller.  Note that if $L \lesssim 1$,
the problem can be tackled simply using the linear theory, as $\mid
\Psi \mid << 1$.

For sources with $L \simeq 1$ or smaller, nothing can be done in the
non-linear regime. The problem is essentially a $(d+1)$-dimensional
one. However, for $L >> 1$, in the above derivative expansion, we see
the leading term has no $y$ dependence. One expects the remaining
terms to be small, and thus we see might hope the non-linearity is not
truly $(d+1)$-dimensional, but rather simply $d$-dimensional. We now
explicitly see how to realize this.

%
\subsection{Orbifold Reduction}
%

We now illustrate the orbifold reduction technique employed in
\cite{Lukas:1998ew,Lukas:1999yn}, the issue being whether one can do
better than perturbation theory for slowly varying sources.  To
proceed we observe that the leading term in the expansion
\eqref{eq:toyexp}, whilst being large, is independent of $y$.
Therefore instead of linearizing the equation about $\Psi(x,y) = 0$ as
above, we try separating a homogeneous piece of the solution from the
inhomogeneous part, as,
\begin{equation}
\Psi(x,y) = H(x) + \phi(x,y)
\end{equation}
where we choose that $\phi(x,0) = 0$ to define the splitting. The aim
is to absorb the leading term of \eqref{eq:toyexp} into $H(x)$.  Then
$\phi(x,y)$ will consist of the remaining terms in the expansion
\eqref{eq:toyexp}, but these are all small. Note that $H(x)$ is a zero
mode solution only when $\rho(x) = 0$. The equation becomes,
\begin{equation}
\lap_d H(x) + (\partial_\mu H)^2 + \lap_d \phi +
2 (\partial_\mu H) (\partial^\mu \phi) + (\partial_\mu \phi)^2 + \partial_y^2 \phi + 
(\partial_y \phi)^2 = 0
\end{equation}
and assuming $\mid \phi \mid << 1$, can be linearized to eliminate
quadratic terms in $\phi$,
\begin{equation}
\lap_d H(x) + (\partial_\mu H)^2 + \lap_d \phi + 2
(\partial_\mu H) (\partial^\mu \phi) + \partial_y^2 \phi  = 0
\end{equation}
Removing such quadratic terms is the crucial step which will reduce
the problem to a $d$-dimensional one.

%
\subsection{`Strong Gravity'}
%

In order to characterize the magnitude of the terms we define,
\begin{equation}
  \mid \partial^p_{\mu} \rho(x) \mid \sim \frac{\rho}{L^p}
\end{equation}
for any integer $p \ge 0$. We characterize,
\begin{equation}
\mid \frac{1}{\lap_d} \rho(x) \mid \sim L^2 \rho = \Phi
\end{equation}
which for large $L$ allows $\Phi \sim O(1)$ even for small $\rho$. We now
assume that we can perform the split so that,
\begin{align}
H(x) & = \left[ \frac{h_0}{\lap_d} + h_1 + h_2 \lap_d + \ldots
\right] \rho(x) \sim
O(\Phi) = O(L^2 \rho)
\notag \\ \notag \\
\phi(x,y) & = \left[ f_1(y) + f_2(y) \lap_d + \ldots \right] \rho(x) \sim O(\rho)
\end{align}
where $h_i$ are constants and $f_i(y)$ are only functions of $y$.
Then we assess the terms above as,
\begin{center}
\begin{tabular}{ccccc}
$\lap_d H$ & $(\partial_\mu H)^2$ & $(\partial_\mu H) (\partial^\mu
\phi)$ & $\partial_y^2 \phi$ & $\lap_d \phi$  
\\ \\
$\frac{1}{L^2} (L^2 \rho) = O(\rho)$ & $(\frac{1}{L} (L^2 \rho))^2 = O(\Phi \rho)$ & $(\frac{1}{L} (L^2 \rho))
(\frac{1}{L} (\rho)) = O(\rho^2)$ & $O(\rho)$ & $O(\frac{1}{L^2} \rho)$
\end{tabular}
\end{center}
and now we see that in fact the cross term, $(\partial_\mu H)
(\partial^\mu \phi)$ is of order $O(\rho^2)$, and can again be
neglected. We are left with the equation,
\begin{equation}
\lap_d H + (\partial_\mu H)^2 + \partial_y^2 \phi = - \lap_d \phi
\label{eq:toyeq}
\end{equation}
where terms on the left hand side are $\sim O(\rho), O(\rho \Phi)$ and
on the right are order $\sim O(\rho / L^2)$. We also see that $H$ and
$\phi$ have decoupled. The boundary condition for the problem is that,
$\phi = 0$ at $y = 0$ and $\partial_y \phi = 0$ at $y = 1$. We may
solve the system by taking some $H(x)$ and then solving the linear
problem,
\begin{equation}
\lap_d \phi(x,y) + \partial_y^2 \phi(x,y) = - D_d[H(x)]
\end{equation}
where $D_d[H(x)] = \lap_d H(x) + (\partial_\mu H)^2$ and provides a
homogeneous source term for $\phi$, of order $O(\rho)$, even when $H \sim
1$. This equation can be solved exactly using a Greens function, and
then derivative expanded as $L >> 1$ as,
\begin{equation}
\phi(x,y) = \left[ \left( y - \frac{1}{2} y^2 \right) + \left(
    \frac{1}{3} y - \frac{1}{6} y^3 + \frac{1}{24} y^4 \right)
\lap_d + O(\lap_d^2) \right] D_d[H(x)]
\end{equation}
Note that whilst we can solve this equation exactly, it is only useful
if $L >> 1$, as otherwise terms in the expansion will not decrease in
magnitude. However for $L \lesssim 1$, $\Phi << 1$ in any case, and
linear theory can be used.

Now we have solved the (d+1)-dimensional problem as $\Psi(x,y) = H(x)
+ \phi(x,y)$, with the source $\rho(x) = \partial_y \Psi \mid_{y =
  0}$,
\begin{equation}
\rho(x) = \left[ 1 + \frac{1}{3} \lap_d + O(\lap_d^2) \right] D_d[H(x)]
\end{equation}
and the non-linear $(d+1)$-dimensional problem is reduced to a
$d$-dimensional one, which was the purpose of the exercise. Note also
that we have not inverted Laplacians, and thus all issues of
$d$-dimensional boundary conditions are implicit in the reduced
equation. Specifying these, one may then invert the problem,
\begin{align}
  D_d[H(x)] & = \left[ 1 + \frac{1}{3} \lap_d + O(\lap_d^2)
  \right]^{-1} \rho(x)
  \notag \\ \notag \\
  & = \rho(x) + \mbox{KK corrections}
\end{align}
where the zero modes are solutions for $\rho(x) = 0$.

%
\subsection{Regimes of Interest}
%

There are then two regimes of interest to us. Firstly the regime where
$\Phi \sim 1$. In this case the sub-leading term in the derivative
expansion $\lap_d \rho(x) \sim O(\rho / L^2)$ is of order $O(\rho^2)$.
Thus in this strong gravity regime, only the leading term in this
derivative expansion is of relevance as we have not calculated the
other $O(\rho^2)$ corrections due to linearizing in $\phi$. However
the homogeneous split has allowed us to solve the non-linear problem
up to $O(\rho^2)$ corrections. This is the case analogous to strong
gravity on the brane for large objects, where effective corrections to
the Einstein equations will be of order $O( l^2 \kappa^4_{d} \rho^2)$,
with $l$ the AdS length, $\kappa_4^2$ the $d$-dimensional
gravitational constant, and $\rho$ the characteristic matter energy
density.  These are extremely small, and not of relevance in the
strong field regime.

The second case is where $\Phi << 1$, the weak gravity analogy. In
this case we must contrast non-linear terms of order $O(\Phi \rho)$
with terms in the derivative expansion, of order $O(\rho / L^{2p})$.
The former is a purely $d$-dimensional correction, whereas the latter
depends on the size of the object compared to the fundamental length
scale of the compactification. Note, in this example we chose units so
that this fundamental scale was one. In the brane case, we are
concerned with the ratio $L / l$, with $l$ the AdS length.  For large
$L$, the non-linear terms in $H(x)$, $\sim O(\Phi \rho)$, will be
large compared to the derivative corrections.  Simple linear theory
would eliminate these non-linear terms in $H(x)$, and therefore this
method allows usual $d$-dimensional non-linearity to be automatically
included. Of course the linear Greens function solutions tell one
about all sizes of object and therefore contain vastly more
information than the above method can yield. However, in cases where
one is simply using the linear theory to derive the bulk metric in the
long wavelength regime, it is far more powerful to use the techniques
here.

Note that for large enough $p$, one expects the derivative expansion
term $O(\rho / L^{2p})$ to be of order $O(\rho^2)$, and then further
terms no longer give meaningful corrections without calculating
non-derivative $O(\rho^2)$ corrections too. The number of terms in the
derivative expansion that are relevant before $O(\rho^2)$ is reached
depends on the exact values of $\rho, L$.  Certainly the first
sub-leading term, $p = 1$, is always important if $\Phi << 1$, having
magnitude $O(\rho^2 / \Phi)$.

%
\section{Non-Linear Metric Decomposition}
%

In following sections we use the orbifold reduction to solve the
non-linear field equations for an orbifold with matter on the positive
tension brane, and a vacuum negative tension brane. The solution will
relate the $(d+1)$-dimensional geometry to the usual $d$-dimensional
non-linear Einstein equations. We take the $(d+1)$-dimensional bulk
metric,
\begin{equation}
ds^2 = \frac{l^2}{z^2} ( g_{\mu\nu}(x,z) dx^\mu dx^\nu + dz^2
) 
\label{eq:metric1}
\end{equation}
with Greek indices taking values over the brane spacetime
$d$-dimensions. If $g_{\mu\nu}(x)$ is the Minkowski metric then this
is simply AdS in Poincare coordinates, supported by a bulk
cosmological constant $\kappa^2_{d+1} \Lambda = - d(d-1)/(2 l^2)$,
solving the bulk Einstein equations,
\begin{equation}
G_{AB} = - \kappa^2_{d+1} \Lambda g^{(d+1)}_{AB}
\end{equation}
where $A,B$ are $(d+1)$-dimensional spacetime indices and
$g^{(d+1)}_{AB}$ is the metric as in \eqref{eq:metric1}. In vacuum,
positive and negative tension orbifold branes can be supported at $z =
z_1, z_2$ respectively, with $z_1 < z_2$. Observers are taken to
reside on the positive tension $\mathbb{Z}_2$ orbifold brane in
addition to localized stress energy.  We leave the negative tension
brane in vacuum simply for convenience.  However, all the methods
outlined in this paper can be applied relaxing this condition. Note
that the linear theory \cite{Garriga:1999yh} already shows that an
observer on the negative tension brane sees a phenomenologically
unacceptable gravity theory, unless the orbifold is stabilized
\cite{Tanaka:2000er}.  The orbifold $\mathbb{Z}_2$ planes and
localized matter $\indT_{\mu\nu}$ are treated in the thin wall
approximation, the planes having tensions $\kappa^2_{d+1} \sigma = \pm
2 (d - 1) / l$.

Following the methods of
\cite{Lalak:1998ti,Lukas:1998fg,Lukas:1998tt,Lukas:1998ew,Lukas:1999yn},
illustrated previously for the scalar field example in section
\ref{sec:scalar}, we decompose $g_{\mu\nu}$ into a homogeneous and
inhomogeneous piece with respect to the coordinate $z$, as
\begin{equation}
g_{\mu\nu}(x,z) = \g0_{\mu\nu}(x) + h_{\mu\nu}(x,z)
\label{eq:metric2}
\end{equation}
There is obviously freedom in the above decomposition. To uniquely
define it, we require that $g_{\mu\nu}(x,z_1) = \g0_{\mu\nu}(x)$ for
some constant $z_1$, which will be the position of the positive
tension brane if the matter has vanishing stress energy trace. Then
the intrinsic metric on this surface $z_1$ will just be $
\g0_{\mu\nu}(x) l^2 / z_1^2$. If the metric $\g0_{\mu\nu}$ is Ricci
flat, the $(d+1)$-dimensional Einstein equations are solved for
vanishing $h_{\mu\nu}$ \cite{Brecher:1999xf}.  These geometric
deformations are then the gravitational zero modes of the orbifold.

As discussed, we are not able to include the radion zero mode in the
background, and have it remain homogeneous, due to the warped geometry
\cite{Charmousis:1999rg}. Thus we only include the homogeneous
gravitons, $\g0_{\mu\nu}(x)$.  Instead we will include the radion by
extending the brane deflection ideas of the linear theory. Thus we do
not perturb the $zz$ metric component as we will include any
perturbative back-reaction of the radion in the non-linear deflection
of the brane.

The procedure we outline below is to find a suitable $\g0_{\mu\nu}(x)$
such that $h_{\mu\nu}(x,z)$ remains small for a low density, long
wavelength matter perturbations on the orbifold brane, even when the
intrinsic geometry is non-linear.  Thus we aim to absorb all the
non-linearity into $\g0_{\mu\nu}(x)$, which effectively shapes the
induced geometry on the brane. One can think of $h_{\mu\nu}$ as the
contribution of the massive \KK modes, which provide only small
corrections to the induced geometry on the brane, but play the
essential role of supporting the localized matter.

As in the scalar field example in section \ref{sec:scalar}, we define
two dimensionless quantities. The first, $\rho$, characterizes the
matter density or curvature scale compared to the AdS energy density.
The second, $L$, compares the length scale associated with the matter,
to the AdS length. For convenience we consider the parameter $\epsilon
= 1 / L^2$ which is small in the relevant large object limit. Formally
we take,
\begin{align}
  \rho & = l^2 \| \indR^{\mu\nu}{}_{\alpha\beta} \|
  \notag \\ \notag \\
  \rho \, \epsilon & = l^4 \| \indlap \indR^{\mu\nu}{}_{\alpha\beta}
  \|
\label{eq:param}
\end{align}
and then expect,
\begin{align}
  \rho \, \epsilon^p \sim O( \| (l^2 \indlap)^p l^2
  \indR^{\mu\nu}{}_{\alpha\beta} \| )
\label{eq:param2}
\end{align}
where $p \ge 0$ and $\| \|$ indicates the maximum absolute value of
the tensor over all space, with $l$, the AdS length, introduced to
make $\rho, \epsilon$ dimensionless. The curvature tensors are formed
from the induced brane metric, as is the Laplacian $\indlap$, and have
indices arranged as they would appear in curvature invariants. In
order to linearize the Einstein equations in the Kaluza-Klein
contribution $h_{\mu\nu}$, we will require that the matter be low
density compared to AdS scales. In addition, in order to use a
derivative expansion, we will require the object to have a large size
or dynamical time compared to the AdS length.  Thus we make the
requirements that,
\begin{equation}
\rho , \, \epsilon << 1
\label{eq:restrictions}
\end{equation}
As mentioned previously, for $\rho << 1$ and $\epsilon \gtrsim 1$, the
problem can be solved simply using standard linear theory.  An
important point is that this restricts attention to non-singular
geometries, as the curvatures must remain bounded, and our
approximation will work only when they are well below the AdS
curvature scale $1 / l^2$. For consistency we take the induced metric,
$\tilde{g}_{\mu\nu}$, to be of order,
\begin{equation}
 \|
(l \partial_\mu)^{2 p} \tilde{g}_{\mu\nu} \| \sim O( \frac{\rho}{\epsilon}
\, \epsilon^p ) \quad \mbox{for} \quad p \ge 1
\end{equation}
Of course, when considering $\| \tilde{g}_{\mu\nu} \|$, the Ricci flat
zero modes must be taken into account. Thus we characterize the
quantity,
\begin{equation}
\Phi = \| \Delta \tilde{g}_{\mu\nu} \| \sim O(\frac{\rho}{\epsilon})
\end{equation}
where we understand $\Delta \tilde{g}_{\mu\nu}$ to be the difference
of the metric from some Ricci flat zero mode background due to the
presence of matter. We note that $\Phi$ will be of order the Newtonian
potential for static field configurations. 

As a technical note, we will also consider the magnitudes in equation
\eqref{eq:param2} to hold, with the tensor indices arranged
differently. This implicitly assumes that the metric is non-degenerate
and non-singular.

%
\section{The Bulk Metric}
\label{sec:metric}
%

In this section, we solve the bulk geometry using a coordinate system
which extends the Randall-Sundrum gauge to the non-linear case
\cite{Garriga:1999yh,Giddings:2000mu}. As in the linear theory, this
coordinate system is such that constant $z$ hyper-surfaces have scalar
induced extrinsic curvature equal to $d / l$. We take a positive
tension brane with localized matter at $z = z_1$, and a vacuum
negative tension brane at $z = z_2$, subject to the restrictions
\eqref{eq:restrictions}, and the further condition that $z_1, z_2 \sim
O(l)$.  The geometry will be consistent only for positive tension
brane matter with a vanishing stress energy trace. In the subsequent
sections we remove this restriction, bending the brane relative to the
surface $z = z_1$, again in analogy with the linear theory, although
now these `radion' deflections may be large, but slowly varying.

Using the metric \eqref{eq:metric1} and zero mode decomposition
\eqref{eq:metric2}, the Einstein equations can be linearized in the
Kaluza-Klein perturbation $h_{\mu\nu}$.  As indicated in the scalar
field example of section \ref{sec:scalar}, we will find that $\|
h_{\mu\nu} \| \sim \| l \dz{h_{\mu\nu}} \| \sim O(\rho)$ and therefore
in linearizing we neglect terms of order $O(h^2 / l^2) \sim O(\rho^2 /
l^2)$, consistent with the low curvature condition. Away from the
branes, the linearized Einstein equations, with critical bulk
cosmological constant, are,
\begin{align}
  G_{zz} - \kappa^2_{d+1} \Lambda g^{(d+1)}_{zz} & = - \hat{R} -
  \frac{d-1}{2 z} \dz{h} + O(h^2 / l^2)
  \notag \\ \notag \\
  G_{z\mu} & = \frac{1}{2} \left( \ograd_{\alpha}
    \dz{h^{\alpha}{}_{\mu}} - \ograd_{\mu} \dz{h} \right) + O(h^2 /
  l^2)
  \notag \\ \notag \\
  G_{\mu\nu} - \kappa^2_{d+1} \Lambda g^{(d+1)}_{\mu\nu} & =
  \hat{G}_{\mu\nu} - \frac{1}{2} ( \ddz{h_{\mu\nu}} - \g0_{\mu\nu}
  \ddz{h} ) + \frac{d-1}{2 z} ( \dz{h_{\mu\nu}} -
  \g0_{\mu\nu} \dz{h} ) + O(h^2 / l^2)
\end{align}
where $\hat{R} = \g0^{\mu\nu} \hat{R}_{\mu\nu}$, and
$\hat{R}_{\mu\nu}, \hat{G}_{\mu\nu}$ are the Ricci and Einstein
curvature of the metric $g_{\mu\nu}(x,z)$ on constant $z$
hyper-surfaces. Indices are raised and lowered with respect to
$\g0_{\mu\nu}(x)$, $h = h^{\alpha}{}_{\alpha}$ and $\ograd$ is the
covariant derivative of this homogeneous mode metric. Thus, we are
using $\g0_{\mu\nu}$ in the same way one uses a background solution in
usual linear perturbation theory. However, here, $\g0_{\mu\nu}$ is not
a solution to these equations in the presence of brane sources.

As $\| h_{\mu\nu} \| \sim O(\rho)$, we may decompose the
$d$-dimensional curvature terms. The Ricci curvature term in $G_{zz}$
becomes
\begin{align}
  \hat{R} & = \g0^{\mu\nu} \hat{R}_{\mu\nu} = \g0^{\alpha\beta}
  \R0_{\alpha\beta} - \olap h + \ograd^{\alpha} \ograd^{\beta}
  h_{\alpha\beta} - \R0^{\alpha}{}_{\beta} h_{\alpha}{}^{\beta} +
  O(h^2 / l^2)
  \notag \\ \notag \\
  & = \R0 - \olap h + \ograd^{\alpha} \ograd^{\beta} h_{\alpha\beta} +
  O(\rho^2 / l^2)
\end{align}
Note that the term $\R0^{\alpha}{}_{\beta} h_{\alpha}{}^{\beta}$ is of
order $O(\rho^2 / l^2)$, as both $l^2 \R0_{\alpha\beta}$ and
$h_{\alpha\beta}$ are of order $O(\rho)$. We now see that the
linearized constraint equations, $G_{zz} - \kappa^2_{d+1} \Lambda
g^{(d+1)}_{zz} = O(\rho^2 / l^2)$, $G_{z\mu} = O(\rho^2 / l^2)$ are
satisfied if,
\begin{align}
  \R0(x) & = 0
  \notag \\ \notag \\
  h^{\alpha}{}_{\alpha}(x,z) & = \ograd_{\alpha}
  h^{\alpha}{}_{\mu}(x,z) = 0
\label{eq:constraints}
\end{align}
This can be recognized as the usual Randall-Sundrum gauge condition.
We have simply imposed this about a homogeneous background
$\g0_{\mu\nu}$, which satisfies the condition $\G0^{\alpha}{}_{\alpha}
= 0$. The remaining Einstein equation becomes,
\begin{equation}
\G0_{\mu\nu} - \frac{1}{2} z^{d-1}
\dz{\Big(} \frac{\dz{h_{\mu\nu}}}{z^{d-1}} \Big) = 
\frac{1}{2} \olap h_{\mu\nu} + O(\rho^2 / l^2)
\label{eq:bulkeqn}
\end{equation}
We emphasize that $\G0_{\mu\nu}$ and $\olap$ are independent of $z$.
Note that $\| \G0_{\mu\nu} \|, \, \| \partial^2_z h_{\mu\nu} \| \sim
O(\rho / l^2)$, and $\olap h_{\mu\nu} \sim O(\rho \, \epsilon / l^2)$.
For slowly varying matter, $\epsilon << 1$, and we construct the
solution as a derivative expansion in $\olap$.  For our brane
configuration, with matter only on the positive tension brane, we
solve this differential equation using the ansatz,
\begin{equation}
h_{\mu\nu}(x,z) = z^{\frac{d}{2}} f(z,\olap) \G0_{\mu\nu}(x)
\end{equation}
which indeed satisfies the constraints \eqref{eq:constraints} if $\R0
= 0$. The transverse condition is simply the contracted Bianchi
identity, and the traceless condition holds as $\R0$ is zero. Then
\eqref{eq:bulkeqn} reduces to the formal `operator' equation,
\begin{equation}
\ddz{f} + \frac{1}{z} \dz{f} + \left( \olap - \frac{d^2}{4 z^2}
\right) f = 2 z^{-\frac{d}{2}}
\end{equation}
which can be solved exactly to give,
\begin{equation}
  f = f_{\rm PI} + A(\olap) \frac{2^{\frac{d}{2}} \left(\frac{d}{2}\right)
    !}{(\olap)^{\frac{d}{4}}} J_{\frac{d}{2}}(\sqrt{\olap} z) - B(\olap)
  \frac{\pi}{2^{\frac{d}{2}} \left(\frac{d}{2} - 1\right) !}
  (\olap)^{\frac{d}{4}} N_{\frac{d}{2}}(\sqrt{\olap} z) 
\label{eq:formalsoln}
\end{equation}
for arbitrary functions $A(\olap), B(\olap)$, which can be expanded
as,
\begin{equation}
A(\olap) = A_0 + A_1 \olap + \frac{A_2}{2 !} (\olap)^2 + \ldots
\end{equation}
and similarly for $B$, ensuring that the whole expression only
contains positive powers of $\olap$. The particular integral is taken
as,
\begin{align}
  \quad f_{\rm PI} & = \left\{ \begin{array}{ll} \frac{2}{\olap} \left[
        z^{-\frac{d}{2}} + \frac{\pi}{2^{\frac{d}{2}}
          \left(\frac{d}{2} - 1\right) !}  {(\olap)^{\frac{d}{4}}}
        N_{\frac{d}{2}}(\sqrt{\olap} z) \right] & d = \mbox{odd}
      \\ \\
      \frac{2}{\olap} \left[ z^{-\frac{d}{2}} +
        \frac{\pi}{2^{\frac{d}{2}} \left(\frac{d}{2} - 1\right) !}
        {(\olap)^{\frac{d}{4}}} \left( N_{\frac{d}{2}}(\sqrt{\olap} z)
          - \frac{1}{\pi} \log(\olap) J_{\frac{d}{2}}(\sqrt{\olap} z)
        \right) \right] & d = \mbox{even}
    \end{array} \right. 
\label{eq:formalsoln2}
\end{align}
which again can be expressed as a derivative expansion in $\olap$ of
Taylor series form. Then for long wavelength matter, defined through
the condition \eqref{eq:restrictions}, we expect leading terms to
dominate the series. By expanding the Bessel functions for
$\sqrt{\olap} z >> 1$ in the particular integral solution
\eqref{eq:formalsoln2}, we see that the $d$-dimensional zero mode no
longer well approximates the geometry, exactly as one predicts for
small objects which should behave in a manifestly $(d+1)$-dimensional
manner. If $\sqrt{\olap} >> 1$, ie. for small objects, $\epsilon
\gtrsim 1$ and $\Phi = \rho / \epsilon << 1$ so linear theory can be
used. However, the case where it is $z$ that is large, means one
cannot study large non-linear objects in the one brane Randall-Sundrum
case, as one must remove the second brane to distances greater than $z
\gtrsim 1 / \sqrt{\olap}$. This essentially shows why the one brane
case is really a $(d+1)$-dimensional problem.

We must now fix the two functions of integration, $A(\olap),
B(\olap)$, using the boundary conditions of the problem. At $z = z_2$,
the position of the negative tension vacuum brane, we require
$\dz{h}_{\mu\nu}(x,z_2) = 0$. The second boundary condition is simply
the requirement that $h_{\mu\nu}(x,z_1) = 0$ at $z = z_1$ by
definition of the metric splitting \eqref{eq:metric2}. Direct
evaluation of the expressions \eqref{eq:formalsoln} and
\eqref{eq:formalsoln2} for these conditions then determine $A, B$.

Using this formal solution, represented by the operator ${\cal L}(z,
\olap)$ below, we may expand $h_{\mu\nu}$ in powers of $\olap$ acting
on $\G0_{\mu\nu}$, the term $(l^2 \olap)^p \, \G0_{\mu\nu}$ being of
order $O(\rho \, \epsilon^p / l^2)$, so that,
\begin{align}
  h_{\mu\nu} & = {\cal L}(z,\olap) \G0_{\mu\nu} = \left[ h_0(z)
    \mathbf{1} + h_1(z) \olap + O(\olap^2) \right] \G0_{\mu\nu}
  \notag \\ \notag \\
  \mbox{with,} \quad h_0(z) & = B_0 - \frac{1}{d-2} \, z^2 + A_0 z^d
  \notag \\ \notag \\
  h_1(z) & = \left\{ \begin{array}{ll} \frac{1}{2 ( d - 2 )} B_0 z^2 -
      \frac{z^4}{2 ( d - 2 ) ( d - 4 )} + \frac{1}{2 ( d + 2 )} A_0
      z^{d+2} + B_1 - \frac{1}{d-2} \, z^2 + A_1 z^d & d = 3, d \ge 5
      \\ \\
      \frac{1}{4} B_0 z^2 + \frac{1}{12} A_0 z^{6} -
      \frac{\frac{3}{4} + \log{2} - \gamma}{8} z^4 + \frac{1}{8}
      (\log{z}) z^4 + B_1 - \frac{1}{2} \, z^2 + A_1 z^4 & d = 4
\end{array} \right.
\notag \\ \notag \\
\mbox{and,} \quad A_0 & = \frac{2}{d (d-2)} \frac{1}{z_2^{d-2}} , \quad B_0
= \frac{z_1^2}{d-2} \left( 1 - \frac{2}{d} \Omega \right) , \quad \Omega = \left(\frac{z_1}{z_2}\right)^{d-2}
\label{eq:derivexp}
\end{align}
where $A_1, B_1$ are trivially calculated from the boundary conditions
but are not shown explicitly here for clarity. Whilst we have expanded
to sub-leading order, $O(l^4 (\olap) \G0_{\mu\nu})$, as discussed
in the scalar field example, in section \ref{sec:scalar}, it is only
the leading term that is relevant for strongly non-linear
configurations with $\Phi \sim O(1)$.  Then the sub-leading correction
to $h_{\mu\nu}$ is of order $O(\rho^2)$, and one must also calculate
the corrections from the linearization of the Einstein equations in
$\dz{h}$ which enter at the same order. The sub-leading corrections are
important in the case when $\Phi < 1$, and the field is weak, but one
wishes to take $d$-dimensional non-linear terms into account, such as
when calculating post-Newtonian corrections.

%
\section{Conformally Invariant Matter}
\label{sec:conf}
%

For a positive tension orbifold brane at constant $z = z_1$, the
induced metric is,
\begin{equation}
\gind_{\mu\nu} = \frac{l^2}{z_1^2} \g0_{\mu\nu}
\label{eq:indmetric1}
\end{equation}
as $h_{\mu\nu} = 0$ at $z = z_1$ by construction. For an orbifold
brane with tension $\sigma$, the Israel matching conditions
\cite{Israel:1966rt} yield the localized stress energy
$\indT_{\mu\nu}$ to be,
\begin{align}
  - \kappa^2_{d+1} \sigma \, \gind_{\mu\nu} + \kappa^2_{d+1}
  \indT_{\mu\nu} & = 2 \left[ K_{\mu\nu} - \gind_{\mu\nu} K \right]_{z
    = z_1}
  \notag \\ \notag \\
  & = - \frac{2 ( d - 1 )}{l} \gind_{\mu\nu} - \frac{l}{z_1}
  \dz{h}_{\mu\nu} + O(\rho^2 / l)
\end{align}
as $\| h_{\mu\nu} \| \sim O(\rho)$ and the trace, $h$, vanishes. The
terms proportional to $\gind_{\mu\nu}$ cancel for the critical value
of tension, $\sigma$, leaving the localized matter,
\begin{equation}
\kappa^2_{d+1} \indT_{\mu\nu} = - \frac{l}{z_1} \left( \partial_z
  {\cal L}(z,\olap) \right)\mid_{z
  = z_1} \G0_{\mu\nu}(x)  
\label{eq:Bulksoln}
\end{equation}
where we have substituted our solution for $h_{\mu\nu}$ from equation
\eqref{eq:formalsoln}.  We see the trace of the stress energy tensor
vanishes, as our construction has $\R0 = 0$ so no trace can be
supported on a brane at constant $z$. A vacuum orbifold brane with
negative tension $-\sigma$, is placed at $z = z_2$, where
$\dz{h}_{\mu\nu}$ is zero.  As the induced metric on the positive
tension brane is simply given by equation \eqref{eq:indmetric1}, the
induced Einstein tensor on the brane $\indG_{\mu\nu} =
\G0_{\mu\nu}(x)$, and the Laplacian of the induced metric is $\indlap
= \olap z_1^2 / l^2$.  Thus the effective Einstein equations for
conformally invariant matter become,
\begin{align}
  \indG_{\mu\nu}(x) & = - \frac{\kappa^2_{(d+1)}}{l}
  \frac{z_1}{\partial_z {\cal L}(z,\frac{l^2}{z_1^2}\indlap)\mid_{z =
      z_1}} \indT_{\mu\nu}(x)
  \notag \\ \notag \\
  & = \frac{1}{1 - \Omega} \kappa^2_{(d)} \indT_{\mu\nu}(x) + O( \epsilon \, \rho / l^2)
  \notag \\ \notag \\
  \mbox{with,} \quad \kappa^2_{(d)} & = \frac{d-2}{2} \frac{1}{l} \,
  \kappa^2_{(d+1)} , \qquad
  \mbox{and,} \quad \indT^{\alpha}{}_{\alpha} = 0
\label{eq:confeqn}
\end{align}
where we have formally summed all `Kaluza-Klein' contributions in the
first line, and given the leading order approximation in the second.
Indices are now raised and lowered with respect to the induced metric
on the brane.  To leading order in the derivative expansion this is
simply $d$-dimensional Einstein gravity for conformal matter.  This
agrees with linear analysis, but of course is not restricted to linear
matter configurations.

In reference \cite{Mukohyama:2001jv}, the sub-leading correction has
been written in the form of higher derivative terms in the action for
the case of 2 branes stabilized by a bulk scalar. This was achieved by
comparing such terms with the linear theory propagator. Here we have
summed all terms in the derivative expansion, and explicitly used the
non-linear theory. It would therefore be interesting to understand
whether these may all be written in such a higher derivative form, as
conjectured in this reference.

The constant $B_0$ in \eqref{eq:derivexp} was derived from the
requirement that $h_{\mu\nu}$ vanish at $z = z_1$. However, we
observes that in general it simply gives a small redefinition of the
homogeneous metric component. Provided $B_0$ is of order $O(1)$, we
might allow $h_{\mu\nu}$ to be non-zero on the brane itself. However,
the fact that $\indG_{\mu\nu} = \G0_{\mu\nu}(x)$ exactly is a direct
consequence of $h_{\mu\nu}$ vanishing on the brane. If $h_{\mu\nu}$
was not zero there, we would have to include this in calculating
$\indG_{\mu\nu}$. We now see the utility of this choice in simplifying
the derivation of the Einstein equations. When considering
non-conformal matter, this is the chief source of complication when
discussing sub-leading terms in the derivative expansion, as in
Appendix B.

%
\section{General Matter and the Leading Order Effective Action}
\label{sec:nonconf}
%

By identifying an extension of the \RS gauge, we have solved
non-linear low energy gravity, providing $\indT^{\alpha}{}_{\alpha}$
vanishes.  The usual $d$-dimensional response to conformal matter is
recovered on a brane at $z = z_1$, with small higher derivative
corrections, which we have formally calculated exactly.  However we
wish to allow the localized matter to have non-vanishing
$\indT^{\alpha}{}_{\alpha}$. In the linearized theory one sees that
placing the brane at non-constant $z$ generates exactly such a trace.
We therefore repeat this procedure non-linearly.  We will find that
the scalar field describing the deflection of the brane relative to
the $z$ coordinates, provides exactly the correct radion degree of
freedom to `replace' the trace of the matter stress energy, as happens
in the linear theory. Note that the vacuum negative tension brane will
remain at $z = z_2$. Of course, using the same methods, one could
extend the analysis to consider matter on the negative tension brane
too.

For general matter we will only consider the leading term in the
derivative expansion. Whilst we formally solve this to all orders in
the previous section for conformally invariant matter, the
non-conformal case is considerably more complicated. For clarity we
only calculate using the leading order term, so the bulk metric is,
\begin{align}
  ds^2 & = \frac{l^2}{z^2} \left[ \left( \g0_{\mu\nu}(x) +
      h_{\mu\nu}(x,z) \right) dx^\mu dx^\nu + dz^2 \right]
  \notag \\ \notag \\
  & = \frac{l^2}{z^2} \left[ \left( \g0_{\mu\nu}(x) + \left( B_0 -
        \frac{1}{d-2} z^2 + A_0 z^d \right) \G0_{\mu\nu}(x) + O(\olap
      \G0) \right) dx^\mu dx^\nu + dz^2 \right]
\label{eq:repeat_metric}
\end{align}
with $A_0, B_0$ as in \eqref{eq:derivexp}, although we note that this
calculation could be extended to include the sub-leading terms in
$\olap \G0$, if required, as discussed in Appendix B.

In order to calculate the matching conditions, we coordinate transform
the metric to a new Gaussian normal system, $\bx^\mu, \bz$, adapted to
the brane, which is located at $\bz = \bz_1$. This is in analogy with
the linear theory, although we now consider large, but slowly varying
brane deflections.  Starting from the metric \eqref{eq:repeat_metric},
we perform the coordinate transformation,
\begin{equation}
z = \bz e^{\phi(\bx)}, \qquad x^\mu = \bx^\mu + \xi^\mu(\bx,\bz)
\end{equation}
where $\phi$ is only a function of $\bx^{\mu}$, and plays the role of
the d-dimensional scalar field parameterizing the deflection. We may
choose the functions $\xi^\mu$ to vanish on the brane for convenience.
The technical derivation of the matching conditions is given in the
Appendix A. Here, we simply outline the main results. The field $\phi$
is treated non-linearly, although again its derivatives are taken to
be small for large, low density objects.  To leading order in the bulk
derivative expansion, we find the induced metric on the brane is
simply a conformal transformation of $\g0_{\mu\nu}$,
\begin{equation}
\gind_{\mu\nu}(\bx) = \frac{l^2}{\bz_1^2}
e^{-2 \phi(x)} \g0_{\mu\nu}(x) + O(\rho)
\label{eq:induced_metric}
\end{equation}
where we note that $x^\mu = \bx^\mu$ on the brane. The extrinsic
curvature of the leading order metric \eqref{eq:repeat_metric} after
coordinate transformation is then calculated at $\bz = \bz_1$, and
following on from that, the brane matching conditions. We eliminate
$\G0_{\mu\nu}$ by tracing these, as in the linear theory, yielding the
equation for the deflection,
\begin{equation}
  \indlap \phi(x) + \frac{d-2}{2} (\indgrad \phi(x))^2 = - \frac{1}{2
  (d-1)} \frac{\kappa^2_{d+1}}{l} \indT^{\alpha}{}_{\alpha}(x) +
  O(\rho^2 / l^2)
\label{eq:phi_repeat}
\end{equation}
which is, in fact, correct to all orders in the derivative expansion,
$\sim O(\epsilon^p \rho / l^2)$ for $p \ge 0$. Thus the radion
receives no `Kaluza-Klein' corrections as in the linear case
\cite{Charmousis:1999rg}.  By comparing the induced stress energy with
the Einstein curvature of the above induced metric, eliminating
$\G0_{\mu\nu}$ we may then derive the effective Einstein equations.
Finally we are able to construct a local, covariant effective action
for the leading order Einstein equations, in terms of a
$d$-dimensional radion scalar field, and the intrinsic metric on the
brane. Then defining the following, as in Appendix A,
\begin{align}
  \Omega[\phi(x)] & = \left( \frac{\bz_1 e^{\phi(x)}}{z_2}
  \right)^{d-2}, \qquad \kappa^2_{d} = \frac{d-2}{2}
  \frac{\kappa^2_{d+1}}{l}
  \notag \\ \notag \\
  {\cal A}_{\mu\nu}[\phi(x)] & = \left( \indgrad_\mu \indgrad_\nu -
    \gind_{\mu\nu} \indlap \right) \phi(x) - (\indgrad_\mu \phi(x))
  (\indgrad_\nu \phi(x)) - \frac{d-3}{2} (\indgrad \phi(x))^2 \gind_{\mu\nu}
\end{align}
allows us to write the effective Einstein and radion equations
\eqref{eq:approx_einstein} and \eqref{eq:phi_repeat}, to order
$O(\max(l^2 \indlap \indT_{\mu\nu}, \rho^2 / l^2))$ as,
\begin{align}
  (1 - \Omega[\phi(x)]) \indG_{\mu\nu}(x) & = \kappa^2_{d}
  \indT_{\mu\nu} - (d-2) \Omega[\phi(x)] {\cal A}_{\mu\nu}[ \phi(x)]
  \notag \\ \notag \\
  \frac{\indR}{2} & = - {\cal A}^{\alpha}{}_{\alpha}[\phi(x)]
\end{align}
which result from the variation of the local action,
\begin{equation}
S = \frac{1}{2 \kappa^2_{d}} \int d^dx \left( \Psi R -
  \omega[\Psi] \frac{(\indgrad_\alpha \Psi)^2}{\Psi}
\right) + \int d^dx {\cal L}_{\mathrm{matter}}[\gind_{\mu\nu}]
\label{eq:action}
\end{equation}
where,
\begin{equation}
\Psi[\phi] = 1 - \Omega[\phi] = 1 - \left(
  \frac{\bz_1}{z_2} e^{\phi(x)} \right)^{d-2}, \qquad \mbox{ and, }
  \quad \omega[\Psi] = \frac{(d-1)}{(d-2)} \frac{\Psi}{1-\Psi}
\end{equation}
and $-\frac{1}{2} \indT_{\mu\nu} \delta \gind^{\mu\nu} = \delta {\cal
  L}_{\mathrm{matter}}[\gind_{\mu\nu}]$, with the matter Lagrangian
having no dependence on the scalar $\phi$. The approximation of terms is fully discussed in the Appendix A. In
addition, we discuss the calculation of sub-leading terms in the
derivative expansion in Appendix B.

We see that this is a general scalar-tensor theory for non-linear
perturbations of $\phi$.  For linear fluctuations we find,
\begin{align}
\omega[\Psi] & = \frac{d-1}{d-2} \left( (\frac{z_2}{\bz_1})^{d-2} - 1
\right) + O(\phi)
\notag \\ \notag \\
& = \frac{3}{2} \left( e^{2 y / l} - 1 \right) , \qquad \mbox{for}
\quad d = 4 
\end{align}
where $y$ is the proper separation between the unperturbed branes.
Thus for linear perturbations we find Brans-Dicke, with coupling as in
\cite{Garriga:1999yh}. Non-linearly the action is not Brans-Dicke. The
same phenomenological gravitational constraint applies to the brane
separation as in the linear theory \cite{Garriga:1999yh}, namely that
$y \ge 4 l$.

The form agrees with the non-linear zero mode action of
\cite{Chiba:2000rr}, which was derived excluding matter and only
including the orbifold zero modes. Strictly speaking, the orbifold
zero modes do not show which conformal metric the matter couples to.
Of course our treatment includes the graviton, matter and radion
together, non-linearly. In \cite{Csaki:1999mp,Goldberger:1999un} the
effective action was again derived through cosmological
considerations, and again agrees, although, as discussed in
\cite{Charmousis:1999rg,Chiba:2000rr}, the radion ansatz used is
incorrect, although does give the correct effective action for
cosmological symmetry.

There has been much work considering non-linear phenomena in
scalar-tensor theory, with a view to observable tests
\cite{Will:2001mx}.  Here we have the explicit form of the non-linear
effective theory, and it will be interesting future work to explore
its non-perturbative implications. One important point remains, namely
the geometric interpretation of the scalar as the radion. If strong
gravity phenomena, either static or dynamical, lead to a large, but
order $O(1)$ $\phi$ being generated, the branes could be forced
together.  This is discussed further in section \ref{sec:star}.

%
\section{Geometric Interpretation of the Radion Scalar Field}
\label{sec:radion}
%

Having a consistent, non-linear realization of the low energy radion,
allows us to formulate its `meaning' geometrically, without reference
to a particular background or coordinate system. Firstly, the positive
tension brane is deflected to a position $z = \bz_1 e^{\phi(\bx)}$
relative to the metric \eqref{eq:metric1} of section \ref{sec:metric},
for general matter, the negative tension brane remaining at $z = z_2$.
The proper distance along the $z$ axis between the two branes is,
\begin{equation}
d(x) = \int_{\bz_1 e^\phi}^{z_2} l \frac{dz}{z} = l
\log(\frac{z_2}{\bz_1}) - l \phi(x) 
\end{equation}
and thus the difference in proper distance due to matter is simply
$\Delta d(x) = - l \phi$. In the linear theory one can define this
radion as determining the proper distance between the branes, relative
to the unperturbed background. Non-linearly, the coordinate system of
section \ref{sec:metric} is Gaussian normal to the negative tension
brane, and thus the $z$ axis itself is a geodesic normal to this
brane. Therefore, we find a geometric interpretation of the radion on
the positive tension brane, at a point $x^\mu$, which relates its
value to the proper length of the normal geodesic to the vacuum
negative tension brane, that intersects the positive brane at $x^\mu$.

For non-conformal matter on the negative tension brane, so that this
too must be deflected relative to the $z$ coordinates, we again expect
to find the radion to be related to the proper distance between the
two branes along the $z$ axis. More generally a line along the $z$
axis can be phrased geometrically as a geodesic, normal to surfaces
with constant induced scalar extrinsic curvature, of value $d / l$,
related to the brane tension.

%
\section{Cosmology}
%

We now make contact with the previous result that some cosmologies can
simply be constructed by considering moving branes in AdS
\cite{Ida:1999ui,Mukohyama:1999qx,Vollick:1999uz,Kraus:1999it,Chamblin:1999ya,Binetruy:1999hy,Bowcock:2000cq}.
This arises in our construction, through $\g0_{\mu\nu}$ being either
Minkowski or Milne spacetime, for such cosmological evolutions. This
implies that $\G0_{\mu\nu}$ vanishes, and with it, all the non-local
correction terms. Thus, as one expects, cosmology is simply a result
of the radion, and \emph{not} the graviton zero modes or
`Kaluza-Klein' corrections.

In order to see this, we write the induced FRW metric in conformal
time, as a conformal transformation of Minkowski spacetime,
\begin{equation}
\gind_{\mu\nu} dx^\mu dx^\nu = a(t)^2 \left( - dt^2 + dx_{(d-1)}^2\right)
\end{equation}
However, we now observe from \eqref{eq:induced_metric} that the metric
$\g0_{\mu\nu}$ is similarly related to $\gind_{\mu\nu}$ by a conformal
transformation,
\begin{equation}
\gind_{\mu\nu} = \left( \frac{l^2}{\bz_1^2} e^{-2 \phi} \right) \g0_{\mu\nu}
\label{eq:ind2}
\end{equation}
to leading order in the derivative expansion. Therefore $\g0_{\mu\nu}$
must be a time dependent conformal transformation of Minkowski space.
The remaining constraint is that $\G0_{\mu\nu}$ is traceless, and
therefore the Ricci scalar curvature of $\g0_{\mu\nu}$ must vanish.
This restricts $\g0_{\mu\nu}$ to simply be Minkowski space, or a Milne
universe, and in both cases $\G0_{\mu\nu}$ vanishes, and the bulk is
simply AdS. We have suppressed the fact that the metric above,
\eqref{eq:ind2}, is only true to leading order in the derivative
expansion. However all the corrections to it are functions of
$\G0_{\mu\nu}$, and so the leading order solution that $\g0_{\mu\nu}$
is Minkowski spacetime, is true to all orders.

Thus we have re-derived that fact that a subset of cosmologies
considered in
\cite{Chamblin:1999ya,Ida:1999ui,Mukohyama:1999qx,Vollick:1999uz,Kraus:1999it,Binetruy:1999hy}
can simply be thought of as motion of a brane in pure AdS, which can
be seen more generally for the AdS-Schwarzschild case, via a
coordinate transformation \cite{Mukohyama:1999wi,Bowcock:2000cq}.
Cosmologies based on AdS-Schwarzschild, when the black hole mass is
not small, cannot be seen in this framework as they are not small
deformations of the static Randall-Sundrum geometry. However, the
machinery above could certainly be applied about an AdS-Schwarzschild
vacuum solution with suitably modified brane tensions, and would
remain largely unchanged, allowing all the generalized cosmologies to
be included.

We now briefly consider the validity of our approximations at late
times. Then the brane scale factor is large, and therefore the
positive tension brane is very near to the boundary of AdS, ie. $z
\simeq 0$, and thus $\phi \rightarrow - \infty$. So $\phi \sim O(1)$
is certainly not true. However, $\grad \phi = O(L \rho l) \sim H \simeq
\sqrt{\rho} l$, with $H$ the Hubble constant, which is still compatible
with $L^2 \rho \sim O(1)$. Thus, whilst $\phi$ goes to negative
infinity, the quantity $z = \bz e^\phi$ simply moves more and more
slowly to zero and the approximation still holds for late time
cosmology. In fact we see that at very late times, when $z \rightarrow
0$, the effective scalar-tensor parameter becomes very large,
indicating usual Einstein gravity is recovered.

%
\section{Large Brane Deflection: Stars, Dynamics and Brane Collision}
\label{sec:star}
%

In this section we consider whether branes can be forced together by
low energy matter. In the previous sections, we have outlined the
non-linear, geometric significance of the radion field. In addition,
we have shown it to take values of order $\sim O(1)$ for strong
gravity configurations. Since phenomenological constraints have shown
the branes may be minimally separated such that $z_2 / \bz_1 \sim
e^4$, it may be possible that for very relativistic stars, or in
dynamical collapse, the branes may actually touch, even though all
curvatures remain low compared to the AdS length. It is instructive to
consider the 4-dimensional linear theory for a static geometry, and
perfect fluid matter, so that,
\begin{equation}
\indgrad^2 \phi(\underline{\bf x}) = - \frac{\kappa^2_4}{6} \indT = \frac{\kappa^2_4}{6}
\left( \rho - 3 P \right)
\end{equation}
Positive values of $\phi$ will deflect the branes together, so we
expect such behavior for $\indT > 0$. However, for usual matter $\indT
< 0$, indicating that the branes in fact deflect apart for
perturbative sources. We might expect this for non-linear sources too.
Of course this intuition is based on the linear theory, and we must
solve the full non-linear equations to confirm it.

We numerically integrate the leading order equations from the action
\eqref{eq:action}, for static spherical symmetry.  Such
configurations are relevant for relativistic stars, such as neutron
stars. We do not present an exhaustive numerical study, but merely
some illustrative examples, the solutions being intended to highlight
the geometric nature of the scalar radion. We use a 4-dimensional
spherically symmetric, static ansatz,
\begin{align}
  ds^2 & = - e^{2 T(r)} dt^2 + e^{2 R(r)} dr^2 + r^2 d\Omega_2^2
  \notag \\ \notag \\
  \phi & = \phi(r)
\end{align}
with $d\Omega^2_2$ the area element on an $S^2$. There have been several
numerical studies of stars in Brans-Dicke and `quadratic' models
\cite{Harada:1997mr,Harada:1998ge,Damour:1993hw,Damour:1996ke,Novak:1998rk}.
These studies all use polytropic equations of state, thought to be
relevant for neutron stars. The effect we wish to illustrate is the
dependence of the sign of the deflection on the sign of $\indT = - \rho +
3 P$. Thus we choose the less realistic incompressible fluid matter,
where $\indT$ changes sign as non-linear effects become important. In
standard GR, the core pressure diverges at the upper mass limit of the
star, and thus $\indT$ becomes arbitrarily negative there.

The incompressible fluid is taken to have density $\rho = \rho_0$.  In
the leading order action, the AdS length $l$, only appears in the
usual lower dimensional Planck constant $\kappa^2_4$. Thus the
solution of the above variables for a particular radius star, may be
scaled in the usual manner to a star of any radius. For convenience,
we choose 4-dimensional Planck units so that $r$ is one unit at the
stellar edge. In these units, the density corresponding to the upper
mass limit in standard GR is $\rho_0 = 8 / 3$, and is not expected to
be very different in this scalar-tensor gravity
\cite{Tsuchida:1998jw}.  Whilst this is not Brans-Dicke, the methods
employed to numerically solve the stellar geometry are essentially
identical.  Initial data at $r = 0$ is provided as $\phi_0 = \phi(0),
P_0 = P(0)$, and the solution is assumed to be regular.  The equations
are Taylor expanded about $r = 0$ to derive conditions for the
remaining variables. These are then integrated using a fourth order
Runge-Kutta algorithm. For a given $\rho_0$, one has a two variable
shooting problem in $\phi_0, P_0$.  One must tune $P_0$ in order to
have $P(1) = 0$, and thus vanishing fluid pressure outside the star.
In addition, $\phi_0$ must be varied to obtain $\phi(\infty) = 0$. In
order to estimate this value, we simply integrate the equations to $r
= 10$. However, in a more detailed study one should solve the vacuum
exterior equations as in
\cite{Harada:1997mr,Harada:1998ge,Damour:1993hw,Damour:1996ke,Novak:1998rk}
and match the fields at the stellar edge. Note that for the
scalar-tensor theory at hand, the scalar must take non-trivial values
when matter is present, and thus one does not have to consider the
issue of spontaneous scalarization as in
\cite{Damour:1993hw,Damour:1996ke,Novak:1998rk}.

A further point is the issue raised in \cite{Deruelle:2001fb}, that
the extrinsic curvature of the brane is discontinuous for an
incompressible fluid. Thus we may not expect \eqref{eq:param2} to be
well defined for $p > 0$.  Whilst this is true, it does not effect our
leading order solution. It will be evident in sub-leading terms in the
derivative expansion, which require $\olap \G0_{\mu\nu}$ to be
evaluated, and appear problematic if $\G0_{\mu\nu}$ is discontinuous
as is the case here. The physical resolution is to assume that on
small scales, the fluid equation of state is softened to remove the
problem. This will have negligible effect on the leading order
solution. The only effect will be in the sub-leading terms, the
contribution being suppressed by powers of $l / L_{\rm soft}$, where
$L_{\rm soft}$ is the length scale of the softening.

\begin{figure}
\centerline{\psfig{file=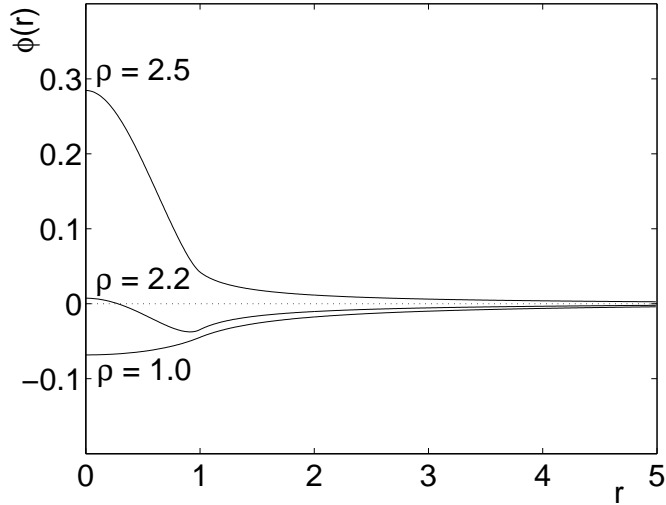,width=3.5in}}
\caption[short]{\figuremode Plots of $\phi$ against $r$ for 3 values of core density, in
  units where the stellar edge is at $r = 1$. For low core density,
  the deflection is negative, the branes moving apart. For high
  density, near to the upper mass limit, $\rho_{\rm max} \simeq
  2.67$, the deflection becomes entirely positive, and the brane
  separation is decreased from the unperturbed case.
\label{fig:phi_vs_r}}
\end{figure}

In figure \ref{fig:phi_vs_r} we present three cases of star, plotting
$\phi$ against the radial coordinate $r$. The brane separation is
chosen to be $4 l$ in proper distance, compatible with the linear
theory phenomenological requirements. However, we find the same
generic behavior for all separations.

We see that the lowest density configuration, with $\rho = 1.0$, does
indeed behave as one expects from the linear theory, the deflection
being negative in $\phi$, and consequently the branes being further
apart at the stellar core, than in the vacuum case. However, as
expected, when the density is increased, $-\rho + 3 P$ becomes
positive in the interior and the deflection starts to become positive,
as for $\rho = 2.2$. For $\rho = 2.5$, even nearer to the upper mass
limit ($\rho_{\rm max} \simeq 2.67$ from standard GR), the value of
the scalar field is entirely positive. Note that for the branes to
touch, $\phi \simeq 4$ is necessary, so all these configurations are
still far from this condition.

\begin{figure}
\centerline{\psfig{file=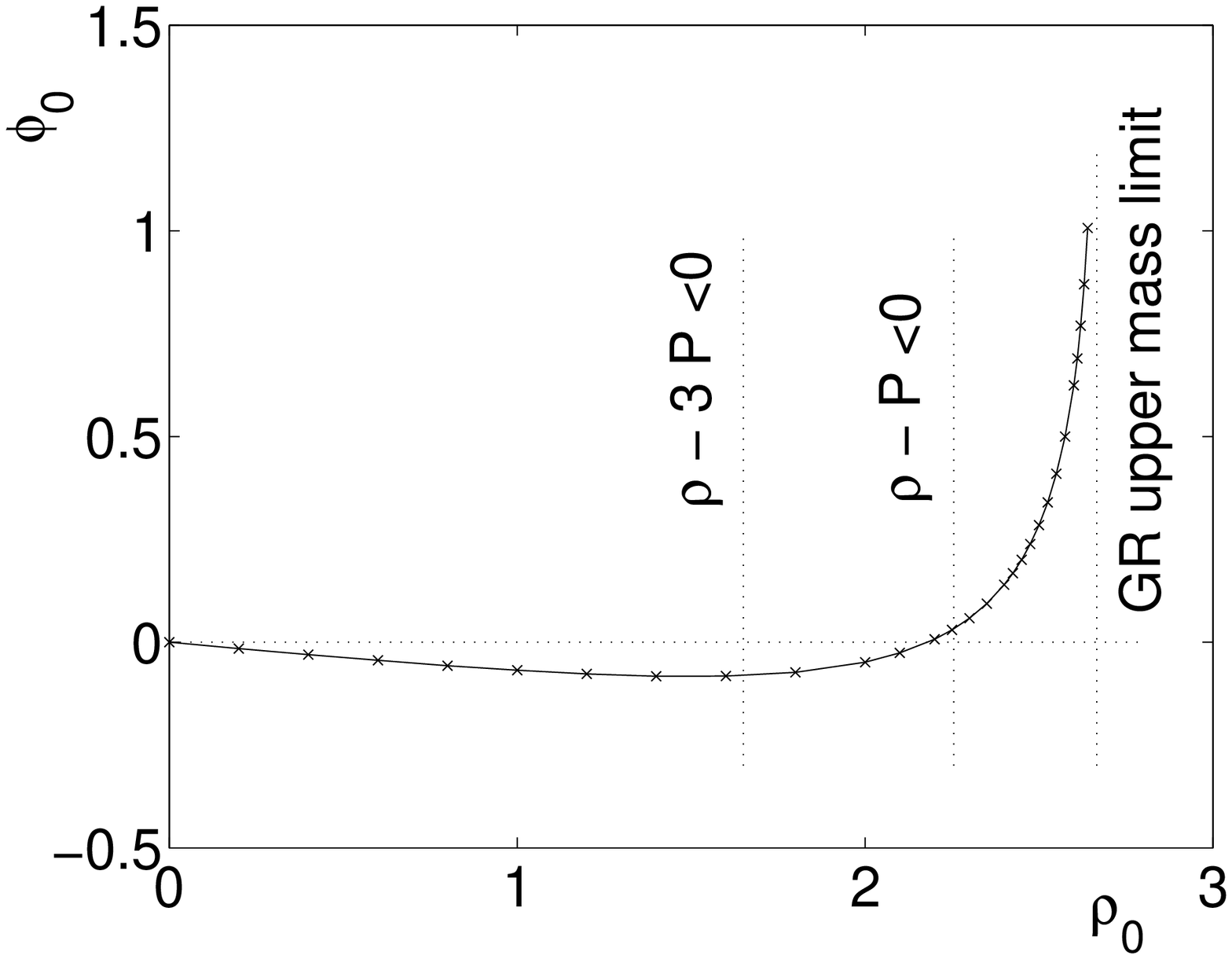,width=3.5in} \hspace{1.0cm} \psfig{file=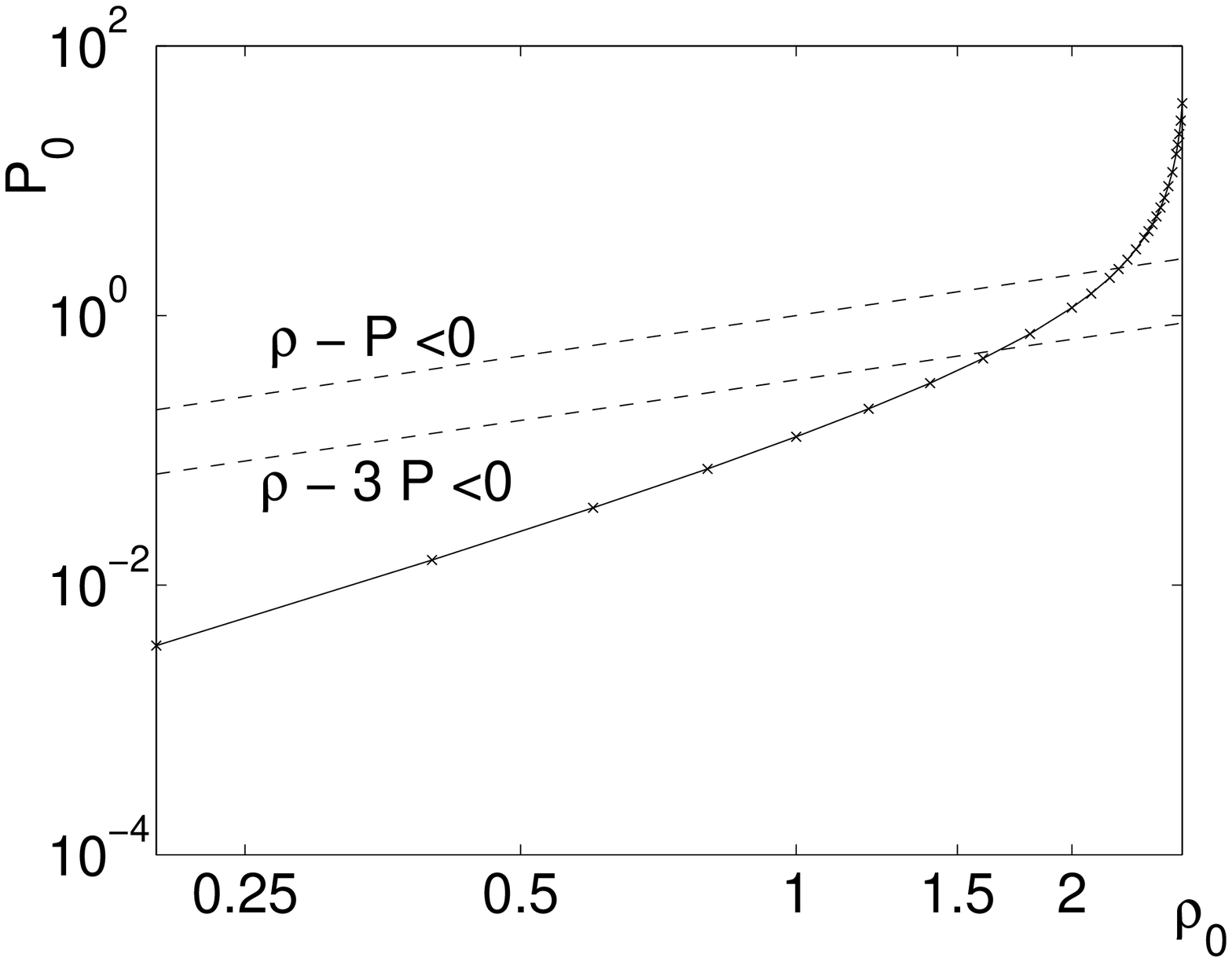,width=3.5in}}
\caption[short]{\figuremode Left-hand frame; Plot of core radion value against stellar
  density, for incompressible fluid stars with fixed angular radius.
  The radion appears to diverge in the positive sense close to the
  upper mass limit. A radion value of $\phi_0 \simeq 4$ would signal
  the branes touch.  Whilst this will appear to happen very close to
  the mass limit, we note that $\phi_0$ has only just become positive
  when the dominant energy condition is violated. Thus, it appears for
  realistic matter, branes will not touch in a static, regular
  geometry. It is then interesting whether realistic matter may cause
  localized brane collisions in dynamical situations. Right-hand
  frame; Plot of core pressure against stellar density. The dominant
  energy condition is plotted, together with $\rho_0 - 3 P_0$, whose
  sign reflects the direction of the brane deflection.
\label{fig:phi_and_P_vs_rho}}
\end{figure}

Figure \ref{fig:phi_and_P_vs_rho} shows the dependence of the core
value of the scalar $\phi$, on the density of the star $\rho_0$. We
see the linear negative deflections, and then the turn-around to
positive deflection for dense stars. The core radion value, $\phi_0$,
apparently diverges in the positive sense as the upper mass limit is
reached, although more detailed numerical analysis would be required
to explore this region.  We mark on this plot the point where $\rho -
3 P$ changes sign. In addition the line $\rho - P = 0$ is shown, being
the point where the dominant energy condition is violated. One sees a
tiny positive core deflection just before this, but nothing
significant. In the same figure, the right-hand plot shows the core
pressure against the density $\rho_0$, with the same lines marked. We
see to gain a deflection of $\phi_0 \simeq 1$, one requires extremely
large core pressures $P \simeq 40$, compared to the core density
$\rho_0 = 2.64$.  The data indicates that arbitrarily large $\phi_0$
can be reached very close to the upper mass limit, although the form
of static matter required to support such a geometry is extremely
unrealistic.

In summary, whilst the scalar radion may become of order $\sim O(1)$
in strong gravity, we find that for static stars, only extreme forms
of matter can support configurations where the branes would meet. It
is then a very interesting question whether the branes may collide in
a dynamical process involving realistic matter. Such dynamics have
been studied for Brans-Dicke
\cite{Shibata:1994qd,Scheel:1995yr,Scheel:1995yn,Harada:1997wt,Balakrishna:1998ek,Kerimo:1998qu}
and it would be interesting to repeat these studies with this
scalar-tensor model and realistic matter, to investigate whether there
are initial configurations that will collapse to give positive $\phi$
large enough to cause a brane collision, in a region which is outside
the formation of an apparent horizon. If such events could occur, it
would indicate that the high energy string theory description of
colliding branes, might be required to understand low energy, low
curvature dynamics in astrophysical processes.

%
\section{Conclusion}
%

We have extended the linearized analysis of the Randall-Sundrum 2
brane compactification to allow the bulk, and induced geometry, to be
calculated in the case of low energy strong gravity, with an
unperturbed, vacuum negative tension brane. The induced geometry must
have low curvature compared to the compactification curvature scale,
and the matter energy density must be slowly varying on scales much
larger than the compactification length. This is the situation
relevant for non-linear gravitation in late time cosmology or
astrophysics.

The radion is inhomogeneous in the transverse coordinate for warped
compactifications. It therefore can not be included as a homogeneous
zero mode, as in usual orbifold reductions. We show how to extend the
coordinate systems used in the linear analysis, to the non-linear, low
energy case. Firstly we choose a coordinate system analogous to the
linear `Randall-Sundrum' gauge. This is Gaussian normal to branes with
conformally invariant matter. We then solve the bulk geometry for
conformally invariant matter on the positive tension brane, using a
formal derivative expansion, which we sum to all orders. General
matter is accommodated by introducing a non-linear `deflection' of the
positive tension brane with respect to this coordinate system. This
deflection is governed by the radion, and is determined to all orders
in the bulk derivative expansion. We are then able to understand the
radion geometrically, as the proper distance between the two orbifold
branes, along a geodesic normal to the vacuum negative tension brane.

Whilst we have applied the method to the 2-brane Randall-Sundrum
geometry, the methods can be extended to any warped compactification
where the radion is not a simple homogeneous mode. Note also, that we
have considered a vacuum negative tension brane, allowing only matter
on the positive brane. This is purely for convenience, and can be
relaxed more generally using the same techniques.

Having solved the bulk geometry, the effective action on the brane is
calculated. To leading order, this yields a scalar-tensor theory with
$\omega \propto \Psi / (1 - \Psi)$, where $\Psi$ is the scalar
depending on the distance between the branes. We have noted that the
radion scalar may take values of order $\sim O(1)$ in a strong gravity
context, allowing the branes to touch or collide when gravitational
fields become strong, and yet curvatures remain small compared to the
compactification scale curvature. We have performed calculations for
static spherical stars using incompressible fluid matter. For very low
density stars, where the response remains linear, the branes are
deflected apart. For densities where the gravitational response
becomes strong, the brane separation may become reduced. We give
evidence that the branes will touch before the stellar upper mass
limit is reached, although the dominant energy condition is certainly
violated. We find that, as indicated in the linear theory, $\rho - 3
P$ must become negative to get an appreciable effect. Whilst this does
occur for incompressible fluid, it is thought not to for physically
reasonable stellar material.  Thus we find unrealistic matter is
required to create touching branes in a static context. It raises the
interesting question of whether realistic matter may allow branes to
locally collide in a dynamic context. If this were the case, high
energy physics may be required to understand low energy astrophysical
processes.

%
\section*{Acknowledgements}
%

It is a pleasure to thank Andrew Tolley and Neil Turok for useful
discussions on this work. The author is supported by a Junior
Research Fellowship at Pembroke College, Cambridge.

\newpage

%
\section*{Appendix A: Non-linear Brane Matching Conditions}
\label{app:non_conf_matching}
%

We now give a detailed derivation of the effective Einstein equations
for general matter, that was outlined in section \ref{sec:nonconf}. We
start by considering the coordinate transformations on the bulk metric
\eqref{eq:repeat_metric},
\begin{equation}
z = \bz e^{\phi(\bx)}, \qquad x^\mu = \bx^\mu + \xi^\mu(\bx,\bz)
\end{equation}
Then preserving the Gaussian normal form of the metric requires that,
\begin{equation}
\dbz{\xi^\mu}(\bx,\bz) = - \bz e^{\phi(\bx)} g^{\mu\alpha}(x,z)
\bar{\partial}_{\alpha} e^{\phi(\bx)} - g^{\mu\alpha}(x,z) g_{\sigma\rho}(x,z) \bar{\partial}_{\alpha}
\xi^\sigma(\bx,\bz) \dbz{\xi^\rho(\bx,\bz)} 
\label{eq:eqnforxi}
\end{equation}
where $\bar{\partial}_{\alpha}, \dbz{}$ are derivatives with respect
to $\bx^{\mu}$ and $\bz$. We see that there is the freedom to take
$\xi^{\mu}$ to vanish on the brane itself, as boundary data for this
differential equation. The metric is then,
\begin{align}
  ds^2 & = \frac{l^2}{\bz^2} \left[ e^{-2 \phi(\bx)}
    \bar{g}_{\mu\nu}(\bx,\bz) d\bx^\mu d\bx^\nu + N^2 d\bz^2 \right]
\notag \\ \notag \\
\bar{g}_{\mu\nu}(\bx,\bz) & = 
      g_{\mu\nu}(x,z) + g_{\alpha(\mu}(x,z) \bar{\partial}_{\mu)}
      \xi^{\alpha}(\bx, \bz) + g_{\alpha\beta}(x,z)
    \bar{\partial}_{\mu} \xi^{\alpha}(\bx,
      \bz) \bar{\partial}_{\nu} \xi^{\beta}(\bx, \bz) + \bz^2 e^{2
    \phi(\bx)} \bar{\partial}_{\mu} \phi(\bx) \bar{\partial}_{\nu} \phi(\bx)    \notag \\ \notag \\
N^2(\bx,\bz) & = 1 + e^{-2
    \phi(\bx)} g_{\alpha\beta}(x,z) (\dbz \xi^{\alpha}(\bx, \bz)) (\dbz
    \xi^{\beta}(\bx, \bz))
\label{eq:trans_metric}
\end{align}
where we temporarily refrain from expanding the metric
$g_{\mu\nu}(x,z)$ in the coordinates $\bx, \bz$.

From the linear theory \cite{Garriga:1999yh,Giddings:2000mu} we might
expect $\phi$ to be given by an expression similar to $\phi \sim (1 /
\lap) \kappa^2_d T$. Therefore $\phi \sim \Phi = O(\rho / \epsilon)$
and of course this cannot be assumed to be small. However, we do
expect that gradients of $\phi$ are small and use this to perform a
gradient expansion of the transformed metric above, as was done for
the bulk geometry in section \ref{sec:metric}. We formalize this with
the assumption
\begin{equation}
\| \bar{\partial}_\alpha e^{\phi(\bz)} \| \sim
O(\frac{\rho}{l \sqrt{\epsilon}})
\, , \qquad  \| \bar{\partial}_\alpha \bar{\partial}_\beta e^{\phi(\bz)} \|
\sim O(\rho / l^2)
\end{equation}
which we will see is consistent with the final result. Now considering
the off diagonal equation \eqref{eq:eqnforxi}, giving $\xi$ in terms
of $\phi$, we see the quadratic term in $\xi$ is an order $O(\rho)$
higher. We find to leading order,
\begin{equation}
  \dbz \xi^{\mu} = - \bz e^{\phi(\bx)} \g0^{\mu\alpha}(\bx)
  \bar{\partial}_{\alpha} e^{\phi(\bx)} \big[ 1 + O(\rho) \big]
\label{eq:approx_xi}
\end{equation}
so that $\dbz \xi^{\mu} \sim O(\frac{\rho}{\sqrt{\epsilon}})$. We have
decomposed $g_{\mu\nu}(x,z)$ into $\g0_{\mu\nu}(x)$ and
$h_{\mu\nu}(x,z)$, and Taylor expanded in $\xi^{\mu} = x^{\mu} -
\bx^{\mu}$. Repeating this for \eqref{eq:trans_metric} we find,
\begin{alignat}{3}
  \bar{g}_{\mu\nu}(\bx,\bz) & = \g0_{\mu\nu}(\bx) + \big[
  h_{\mu\nu}(\bx, \bz e^{\phi(\bx)}) + \xi^\alpha(\bx, \bz)
  \bar{\partial}_{\alpha} \g0_{\mu\nu}(\bx) + \g0_{\alpha(\mu}&&(\bx)
  \bar{\partial}_{\nu)} \xi^{\alpha}(\bx, \bz)
  \notag \\
  & \qquad O(1) \qquad \qquad \qquad \qquad O(\rho) && + \bz^2 (
  \bar{\partial}_{\mu} e^{\phi(\bx)} ) ( \bar{\partial}_{\nu}
  e^{\phi(\bx)} ) \big] + O(\rho^2)
  \notag \\ \notag \\
  N^2 & = 1 + e^{- 2 \phi(\bx)} \g0_{\alpha\beta}(\bx) \dbz
  \xi^{\alpha}(\bx, \bz) \dbz \xi^{\beta}(\bx, \bz) + O(\rho^2) &&
  \notag \\
  & \quad O(1) \qquad \qquad \qquad O(\rho)
\label{eq:approx_trans_metric} 
\end{alignat}
Whilst we may Taylor expand $h_{\mu\nu}(\bx + \xi, \bz e^{\phi})$ in
$\xi$, as $\xi$ is small $\sim O(\frac{\rho l}{\sqrt{\epsilon}})$,
$\phi$ is not small and may be of order one. However, we know the
explicit $z$ dependence of $h_{\mu\nu}$ for each order in the
derivative expansion and therefore we may simply evaluate $h_{\mu\nu}$
at $z = \bz e^\phi$, rather than having to perform a Taylor expansion.
Thus,
\begin{equation}
h_{\mu\nu}(\bx, \bz e^{\phi(\bx)}) = \left( B_0 -
  \frac{1}{d-2} \bz^2 e^{2 \phi(\bx)} + A_0 \bz^d e^{d \phi(\bx)}
  \right) \G0_{\mu\nu}(\bx) + O(l^4 \olap \G0_{\mu\nu})
\label{eq:htrans}
\end{equation}
As in the conformal matter case, $h_{\mu\nu}$ will not contribute to
leading order in the induced metric, but will do in the localized
stress energy. The induced metric on the brane at $\bz = \bz_1$ is
simply,
\begin{equation}
\gind_{\mu\nu}(\bx) = \frac{l^2}{\bz_1^2} 
e^{-2 \phi(\bx)} \bar{g}_{\mu\nu}(\bx,\bz_1) = \frac{l^2}{\bz_1^2}
e^{-2 \phi(x)} \g0_{\mu\nu}(x) + O(\rho)
\label{eq:induced}
\end{equation}
where we use the fact that $\bx^\alpha = x^\alpha$ on the brane. The
$O(\rho)$ terms, which involve $h_{\mu\nu}$, are not relevant if one
is only working to leading order in the derivative expansion, and
their absence makes the following computations considerably simpler.
 
We calculate the stress energy, $\kappa^2_{d+1} \indT_{\mu\nu}(\bx) =
\kappa^2_{d+1} \sigma \gind_{\mu\nu} + 2 \left[ K_{\mu\nu} - K
  \gind_{\mu\nu} \right]_{\bz = \bz_1}$, as,
\begin{align}
  \kappa^2_{d+1} \indT_{\mu\nu}(x) & = -
    \frac{l}{\bz_1} e^{-2 \phi(x)} ( \dbz{h}_{\mu\nu}(\bx,\bz
  e^{\phi(\bx)}) )\mid_{\bz = \bz_1} + 2 l {\cal
    A}_{\mu\nu}[\phi(x)] + O(\rho^2 / l)
\notag \\ \notag \\
{\cal A}_{\mu\nu}[\phi(x)] & =  
    \left( \indgrad_\mu \indgrad_\nu - \gind_{\mu\nu} \indlap \right)
  \phi(x) - (\indgrad_\mu \phi(x)) (\indgrad_\nu \phi(x)) -
    \frac{d-3}{2} (\indgrad \phi(x))^2 \gind_{\mu\nu}
\label{eq:induced_matter}
\end{align}
where $\indgrad$ is the covariant derivative of the induced brane
metric $\gind_{\mu\nu}(x)$ and,
\begin{equation}
\dbz{h}_{\mu\nu}(\bx,\bz
  e^{\phi(\bx)}) )\mid_{\bz = \bz_1} = \left( -
  \frac{2}{d-2} \bz_1 e^{2 \phi(\bx)} + A_0 \, d \, \bz_1^{d-1} e^{d \phi(\bx)}
  \right) \G0_{\mu\nu}(\bx) + O(l^3 \olap \G0)
\end{equation}
As in the linear analysis, the trace of \eqref{eq:induced_matter}
determines $\phi(x)$ as,
\begin{equation}
\gind^{\mu\nu}(\bx) ( \dbz{h}_{\mu\nu}(\bx,\bz
  e^{\phi(\bx)}) ) 
= \frac{\bz^2}{l^2}
e^{2 \phi(\bx)} \dbz{ \left(\g0^{\mu\nu}(\bx) h_{\mu\nu}(\bx,\bz
  e^{\phi(\bx)}) \right) } + O(\rho^2)
= O(\rho^2)
\end{equation}
and $h_{\mu\nu}$ is traceless with respect to $\g0_{\mu\nu}$. This
then determines the location of the brane in the $z$ coordinate system
by specifying that the radion $\phi$ must obey,
\begin{equation}
  \indlap \phi(x) + \frac{d-2}{2} (\indgrad \phi(x))^2 = - \frac{1}{2
  (d-1)} \frac{\kappa^2_{d+1}}{l} \indT^{\alpha}{}_{\alpha}(x) +
  O(\rho^2 / l^2)
\label{eq:phi}
\end{equation}
This result is interesting as it is exact to all orders in the
derivative expansion.  From \eqref{eq:induced} the induced metric is
simply a conformal transformation of the zero mode metric
$\g0_{\mu\nu}$ by $e^{2 \phi(x)}$, to leading order, resulting in an
induced Einstein tensor,
\begin{equation}
  \indG_{\mu\nu}(x) = 
      \G0_{\mu\nu}(x) + (d-2) {\cal A}_{\mu\nu}[\phi(x)] + O(\max(l^2 \olap
    \G0_{\mu\nu}, \rho^2 / l^2)
\label{eq:einstein_tensor}
\end{equation}
where $\olap \G0_{\mu\nu}$ terms appear if one works to higher order
in the derivative expansion including the terms $O(\rho)$ in
\eqref{eq:induced}. All that remains is to solve
\eqref{eq:induced_matter} for $\G0_{\mu\nu}$, and substitute
into \eqref{eq:einstein_tensor} to derive the induced Einstein
equations on the brane, to leading order in the derivative expansion,
\begin{align}
  \indG_{\mu\nu}(x) = \kappa^2_{d} \frac{1}{1 - \Omega[\phi(x)]} &
  \indT_{\mu\nu}(x) - (d-2) \frac{\Omega[\phi(x)]}{1 -
    \Omega[\phi(x)]} {\cal A}_{\mu\nu}[\phi(x)] + O(\max(l^2 \indlap
  \kappa^2_d \indT_{\mu\nu}, \rho^2 / l^2))
  \notag \\ \notag \\
  \Omega[\phi(x)] & = \left( \frac{\bz_1 e^{\phi(x)}}{z_2}
  \right)^{d-2}, \qquad \kappa^2_{d} = \frac{d-2}{2}
  \frac{\kappa^2_{d+1}}{l}
\label{eq:approx_einstein}
\end{align}
which together with the equation \eqref{eq:phi} determining $\phi(x)$,
fully specifies the induced Einstein equations, up to corrections
$O(l^2 \olap \G0_{\mu\nu})$ from sub-leading `Kaluza-Klein' terms.
Tracing these equations, and using the equation for $\phi$, one
obtains,
\begin{equation}
\indR = - \frac{\kappa^2_{d+1}}{l} \indT
\end{equation}
which is true to all orders in the derivative expansion, the
corrections being traceless. 

The sub-leading terms have a complicated form when written in the
induced metric Laplacian, rather than $\olap$, which is the Laplacian
of $\g0_{\mu\nu}$.  However, of course these terms can be evaluated if
required, and are briefly discussed in the following Appendix B.

%
\section*{Appendix B: Sub-leading Non-Local Terms}
\label{app:subleading}
%

We now consider the form that sub-leading non-local terms will take in
the effective Einstein equations.  Note that such corrections are only
of interest when $\Phi << 1$, as otherwise they are of order
$O(\rho^2)$, comparable with the other set of bulk corrections we have
ignored.

In the conformally invariant case we have formally summed the
derivative expansion \eqref{eq:derivexp}. This is considerably more
complicated to do in the non-conformal case, the primary reason being
that $h_{\mu\nu}$ no longer vanishes on the brane, when it is not
located at constant $z$ as in the conformal case. Whilst the induced
metric in the conformal case is simply $\gind_{\mu\nu} = ( l^2 / z_1^2
) \g0_{\mu\nu}$, in the non-conformal case,
\begin{equation}
\gind_{\mu\nu}(\bx) = \frac{l^2}{\bz_1^2}
e^{-2 \phi(x)} \left( \g0_{\mu\nu}(x) + \left[ 
  h_{\mu\nu}(x, \bz_1 e^{\phi(x)})  + \bz_1^2 (
  \partial_{\mu} e^{\phi(x)} ) ( \partial_{\nu}
  e^{\phi(x)} ) \right] \right) + O(\rho^2)
\end{equation}
which is presented above in equation \eqref{eq:induced} only to
leading order as sub-leading terms were not considered. One can rewrite
this as,
\begin{equation}
\frac{\bz_1^2}{l^2} e^{+2 \phi(x)} \gind_{\mu\nu}(x) = 
  \g0_{\mu\nu}(x) + \left[ 
  h_{\mu\nu}(x, \bz_1 e^{\phi(x)})  + \bz_1^2 (
  \ograd_{\mu} e^{\phi(x)} ) ( \ograd_{\nu}
  e^{\phi(x)} ) \right] + O(\rho^2)
\label{eq:indmet2} 
\end{equation}
and then calculate the Einstein curvature of both sides perturbatively
using $\gind_{\mu\nu}$ as the background metric on the left-hand side,
and $\g0_{\mu\nu}$ on the right, giving,
\begin{equation}
  \indG_{\mu\nu}(x) - (d-2) {\cal A}_{\mu\nu}[\phi(x)] = 
      \G0_{\mu\nu}(x) - \frac{1}{2} \olap h_{\mu\nu}(x, \bz
      e^{\phi(x)}) + O(\rho^2 / l^2)
\label{eq:einstein_actual}
\end{equation}
where the term $\bz_1^2 ( \ograd_{\mu} e^{\phi} ) ( \ograd_{\nu}
e^{\phi} )$ from the metric perturbation on the right hand side of
\eqref{eq:indmet2} in fact contributes only to $O(\rho^2 / l^2)$. A subtlety
in this calculation is that $\ograd^\alpha h_{\alpha\mu}(x,\bz
e^\phi)$ is not zero due to the implicit $e^\phi$
dependence. However, one finds that,
\begin{equation}
\ograd^\alpha h_{\alpha\mu}(x,\bz e^\phi) = \left( \ograd^{\alpha}\phi \right)
\bz \partial_{\bz} h_{\alpha\mu}(x,\bz e^\phi) \sim O((L \rho)
(\frac{\rho}{l L}) ) = O(\rho^2 / l)
\end{equation}
and so these terms can be ignored. Indeed, if $\Phi \sim 1$, then $\|
\phi \| \sim 1$ but the first sub-leading term is of order $O(\rho^2)$
and is not useful without calculating the other $O(\rho^2)$
corrections. If $\Phi << 1$, then similarly $\| \phi \| << 1$. One finds
that whilst it is important to consider the $\phi$ dependence in the
leading term of the derivative expansion of $h_{\mu\nu}$, which
results in the recovery of the scalar-tensor behavior, for the
sub-leading terms, $\phi$ corrections are in fact negligible.
Remembering we are considering the case $\| \phi \| << 1$, consider the two
contributions,
\begin{align}
  \phi \G0_{\mu\nu} & \sim (L^2 \rho) (\rho / l^2)
  \notag \\ \notag \\
  \phi \olap \G0_{\mu\nu} & \sim (L^2 \rho) (\frac{1}{L^2} ) (\rho /
  l^2) = O(\rho^2 / l^2)
\end{align}
Whilst the first term, which could be found in the leading term of the
expansion of $h_{\mu\nu}$ is still relevant in the Einstein equations,
the second term, found in the first sub-leading correction, is not as
it is of order $O(\rho^2 / l^2)$.

In order to write the effective Einstein equations as done in equation
\eqref{eq:approx_einstein}, one must eliminate $\G0_{\mu\nu}$ between
\eqref{eq:induced_matter} and \eqref{eq:einstein_actual}. To leading
order in the derivative expansion this is trivial as
\eqref{eq:einstein_actual} is inverted to solve for $\G0_{\mu\nu}$
simply by rearrangement. However to sub-leading order this is a
slightly more tricky procedure in general, as one must invert the
derivative expansion. Of course this can be done if such terms are
required, although we refrain from calculating them here. The final
step is then to convert $\G0_{\mu\nu}$ to $\indG_{\mu\nu}$ and $\olap$
to $\indlap$.

%
\newpage 
%

\newcommand{\href}[1]{}%
\newcommand{\dhref}[1]{}%
\newenvironment{hpabstract}{%
  \renewcommand{\baselinestretch}{0.2}
  \begin{footnotesize}%
}{\end{footnotesize}}%
\newcommand{\hpeprint}[1]{%
  \href{http://arXiv.org/abs/#1}{\texttt{#1}}}%
\newcommand{\hpspires}[1]{%
  \dhref{http://www.slac.stanford.edu/spires/find/hep/www?#1}{\ (spires)}}%

%
\end{document}